# Effect of electron exchange on atomic ionization in a strong electric field

M. Ya. Amusia[1][1,2]

[1] *Racah Institute of physics, The Hebrew University, 91904 Jerusalem, Israel*
[2] *Ioffe Physico-Technical Institute, 194021 St. Petersburg, Russia*

**Key words** Hartree-Fock equation, wave-functions asymptotic, strong field ionization

**PACS** 03.65.Ca, 03.65.Ge, 03.65.Nk, 03.65.Xp

**Abstract**
Hartree-Fock atom in a strong electric static field is considered. It is demonstrated that exchange between outer and inner electrons, taken into account by the so-called Fock term affects strongly the long-range behavior of the inner electron wave function. As a result, it dramatically increases its probability to be ionized.

A simple model is analyzed demonstrating that the decay probability, compared to the case of a local (Hartree) atomic potential, increases by many orders of magnitude. As a result of such increase, the ratio of inner to outer electrons ionization probability became not too small.

It is essential that the effect of exchange upon probability of inner electron ionization by strong electric field is proportional to the square of the number of outer electrons. It signals that in clusters the inner electron ionization by strong field, the very fact of which is manifested by e.g. high energy quanta emission, has to be essentially increased as compared to this process in gaseous atomic objects.

**1**. The aim of this Letter is to describe a mechanism that may be responsible for inner-shell ionization of atoms under the action of strong electric or low frequency laser fields. It was found early in the middle of eighties that the action of the strong low-frequency laser field upon an atom leads with relatively high probability to elimination of several electrons [1]. Some information appeared at that time [2], unconfirmed however by other investigations that in interaction of an eximer laser with Xe emitted photons with energies up to 500 were detected. Estimation has clearly demonstrated that the probability observed is by orders of magnitude higher than results of perturbation theory in the number of absorbed photons.

The way out of this controversy was proposed in [3], where "atomic antenna" mechanism was suggested. The main idea of this mechanism is that the first electron eliminated by the field starts to oscillate in this field, acquiring so-called "pondermotive energy" $E_p$. The remarkable feature of this energy is that it is proportional to the laser intensity W and inversely proportional to the square of its frequency $\omega$. For frequencies of about 1 eV $(1/27.2 \text{ at.un.})^2$ and $W = 10^{16}$ Watts/cm$^2$ (equal to 1 at. un.) $E_p$ is already as big as about 30 keV. An electron oscillating with such energy can collide with the parent ion eliminating other electros, including the inner ones. Immediately after "atomic antenna" was suggested, an idea appeared that such mechanism can generate X-ray as well [4]. About ten years after publication of [3], this mechanism was re-suggested in [5], and became the bases for understanding of ionization atoms in strong electric fields.

At the end of eighties, I start to question the generally accepted method of estimation of the inner electron direct ionization probability. Namely, I came to the conclusion that the asymptotic of inner electron wave functions was estimated incorrectly. From direct calculations we knew and have analytic confirmations (see also [6]) that when exchange between outer and inner electrons (it is in Hartree-Fock approximation) is taken into account the inner electron wave function asymptotic acquire admixture of the outer electron.

---
[1] E-mail: amusia@vms.huji.ac.il
[2] The atomic system of units is used in this paper: electron mass *m*, charge *e* and Planks constant are equal to 1.

On the ground of this asymptotic behavior it was demonstrated that inner shell ionization in a strong electric field is by many orders of magnitude bigger than estimated in the frame of an ordinary one-electron approximation [7]. The reaction of colleagues, first of all Prof. U. Fano, to this idea was lukewarm. This is why a paper on this subject was never written, although I did not found any defects in own argumentation. Perhaps, however, I would never present it as a paper if not recently run across the e-prints [8, 9]. This was the big straw that pushed me at last in the right, as I believe, direction.

**2.** As is well known, the Hartree-Fock equation for an atom in electric homogeneous field $\vec{E}\vec{r}$ field look as follows [10]:

$$\left[ -\frac{\Delta}{2} - \frac{Z}{r} + \vec{E}\vec{r} + \sum_{k=1}^{N} \int \rho_k(x') \frac{dx'}{|\vec{r}'-\vec{r}|} \right] \varphi_j(x) - \sum_{k=1}^{N} \int \varphi_k^*(x') \frac{dx'}{|\vec{r}'-\vec{r}|} \varphi_j(x') \varphi_k(x) \equiv$$
$$\equiv \left[ H_{HF}(x) + \vec{E}\vec{r} \right] \varphi_j(x) - \sum_{k=1}^{N} \int \varphi_k^*(x') \frac{dx'}{|\vec{r}'-\vec{r}|} \varphi_j(x') \varphi_k(x) = E_j \varphi_j(x) \quad . \quad (1)$$

Here $x = \vec{r}, \vec{s}$ is the coordinate and spin projection, $\rho_k(x) \equiv |\varphi_k(x)|^2$ is the one-electron $k$ state density. The total electron $\rho(x)$ density is given by $\rho(x) = \sum_{k=1}^{N} \rho_k(x)$ It is seen from (1) that at large distances the effective potential behaves as $(-Z + N - 1)/r + \vec{E}\vec{r}$.

The contribution of the second term in the integrand cannot be presented as an action of some normal potential $W(r)$ upon $\varphi_j(x)$. On the contrary, the action described by this term is non-local, connecting points $x$ and $x'$, over which the integration is performed.

**3.** The exchange term modifies essentially the long distance behavior of the inner electron wave function. The exchange with the outer electron leads to additional zeroes in the inner electron wave functions and to big increase of the asymptotic value that determines the probability of ionization under the action of static field.

It is well known (see e.g. [11]) that each discrete level in an attractive spherically – symmetric potential is characterized by the following set of quantum numbers: principal $n$, radial $n_r = n - l - 1$, angular momentum $l$ and its projection $m$, and spin projection $s$. It is known that the number of zeroes (or nodes) of the wave function $\varphi_{nlms}(x)$ is equal to $n_r$ and therefore the wave function of the lowest in energy state with $n = 1$, $1s$, has no zeroes. The asymptotic are determined by the binding energy of the level

$$\varphi_{nl}(\vec{r})\big|_{r\to\infty} \approx \alpha_{nl}^{3/2} (\alpha_n r)^{\frac{1}{\alpha_{nl}}-1} e^{-\alpha_{nl} r}, \quad (2)$$

where $\alpha_{nl} = \sqrt{2|E_{nl}|}$ and $E_{nl}$ is the binding energy of the level $nl$.

Solutions of HF equations have extra zeroes, even for the lowest 1s level and the asymptotic of the wave function is determined, contrary to (2), not by $E_{nl}$, but instead by exponent with smallest in absolute value binding energies, if states with higher principal quantum numbers than $n$ are occupied. The number of zeroes is not determined by the radial quantum number $n_r$ of the considered level, but mainly by $n_r$ of the outermost particle.

The general proof of the statement about number of zeroes one can find in [6]. The extra zeroes are located at big distances. Having this in mind, let us consider the asymptotic of the one-particle HF wave function. Let us take for simplicity a two-level "atom", with one inner $i$ and the other outer $o$ electron and consider the equation for the inner state with wave function $\varphi_i(x)$ in order to see how it is modified by the exchange with the outer electron. Instead of (1) we have

$$\left[ -\frac{\Delta}{2} - \frac{Z}{r} + \vec{E}\vec{r} + \int \rho(x') \frac{1}{|\vec{r}'-\vec{r}|} dx' \right] \varphi_i(x) - \int \varphi_o^*(x') \frac{dx'}{|\vec{r}'-\vec{r}|} \varphi_i(x') \varphi_o(x) = -|E_i| \varphi_i(x). \quad (3)$$

where $\rho(x)$ is the total electron wave function. If the last term in the left hand side of (3) is neglected, all the rest decreases according to (2).

At large distances the exchange term $\Re(r)$ behaves as

$$\Re(r)\big|_{r\to\infty} = \int \varphi_o^*(x') \frac{dx'}{|\vec{r}'-\vec{r}|} \varphi_i(x') \varphi_o(x) \bigg|_{r\to\infty} \approx \frac{1}{r^2} \varphi_o(x) \int \varphi_o^*(x')(\vec{r}'\vec{n}) \varphi_i(x') dx' \equiv \frac{C_{n_o}}{r^2} \varphi_o(x). \quad (4)$$

where $\vec{n}$ is the unit vector in the direction $\vec{r}$ and index $n$ stands for the principal quantum number of outer electron.

It is evident from (4) that located inside the "atom" the wave function $\varphi_i(\vec{r})$ acquires an admixture of $\varphi_o(x)$ with much bigger radius of the outer electron. The wave function of an outer electron $\varphi_o(x)$ has more zeroes than has the inner if exchange terms are not taken into account.

If consider for concreteness $i=1s$ then $\varphi_{1s}(\vec{r}) \approx \alpha_{1s}^{3/2} e^{-\alpha_{1s}r}$, where $\alpha_{1s} = \sqrt{2|E_{1s}|}$. As it follows from (4), $s$-state can be mixed with only $p$-states. Substituting a superposition of asymptotes into (3), we obtain using (4) the following expression for asymptotic of $\varphi_i(\vec{r})$:

$$\varphi_i(\vec{r})\big|_{r\to\infty} \approx \alpha_i^{3/2} (\alpha_i r)^{\frac{1}{\alpha_i}-1} e^{-\alpha_i r} + \frac{C_{n_o}}{(\alpha_i r)^2} \alpha_{n_o l}^{3/2} (\alpha_{n_o l} r)^{\frac{1}{\alpha_{n_o l}}-1} e^{-\alpha_{nl}r}. \quad (5)$$

If $\alpha_i$ is considerably bigger than $\alpha_{n_o}$, i.e. if the energy levels are well separated, the first term in (5) can be neglected leading to

$$\varphi_i(\vec{r})\big|_{r\to\infty} \approx \frac{C_{n_o}}{(\alpha_i r)^2} \alpha_{n_o l}^{3/2} (\alpha_{n_o l} r)^{\frac{1}{\alpha_{n_o l}}-1} e^{-\alpha_{nl}r}, \quad (6)$$

thus completely modifying the asymptotic as compared to the case when exchange is neglected. Note that for pure hydrogen field $\alpha_{nl} = 1/n$ and $E_n = -1/2n^2$.

We see that the asymptotic of any one-electron HF occupied state wave function $\varphi_{1s}(\vec{r}) \approx \alpha_i^{3/2} \exp(-\sqrt{2|E_i|}r)$ is determined not by the state's binding energy $E_i$ but can be much bigger, $\sim \exp(-\sqrt{2|E_{min}|}r)$, where $E_{min}$ is the energy of the outermost particle. If there are several outer levels, the effect of exchange is determined by the following expression

$$\varphi_i(\vec{r})|_{r\to\infty} \approx \sum_{\text{All outer } n_o} \frac{C_{n_o}}{(\alpha_i r)^2} \alpha_{n_o l}^{3/2} (\alpha_{n_o l} r)^{\frac{1}{\alpha_{n_o l}}-1} e^{-\alpha_{n_o l} r}, \quad (7)$$

that for $N_o$ outer electrons enhances the exchange influence by this same factor. The role of exchange contribution can be achieved by exciting the outer electrons to states with smaller than minimal in the atomic ground state energies.

The alteration of the one-particle wave function profoundly increases the probability of ionization of the inner levels by a strong laser fields.

Concrete numeric calculations have demonstrated that the exchange increases the magnitude of the 1s wave function in Ar at $r=1$ by a factor as big as $10^{17}$!

**4.** Let us show how long-tail corrections that appear in inner one-electron wave functions due to exchange with outer electrons modify the probability of their elimination from an atom by a strong electric field, of which a concrete example can serve a high intensity (about $10^{18-20}$ and higher Watts/cm$^2$) and low frequency ($\omega \ll I$) laser beam, where $I$ is the atomic ionization potential. The combination of static external and atomic field is depicted in Fig. 1. Let us for simplicity treat a two-level atom considered in section 3. Here we will follow [7].

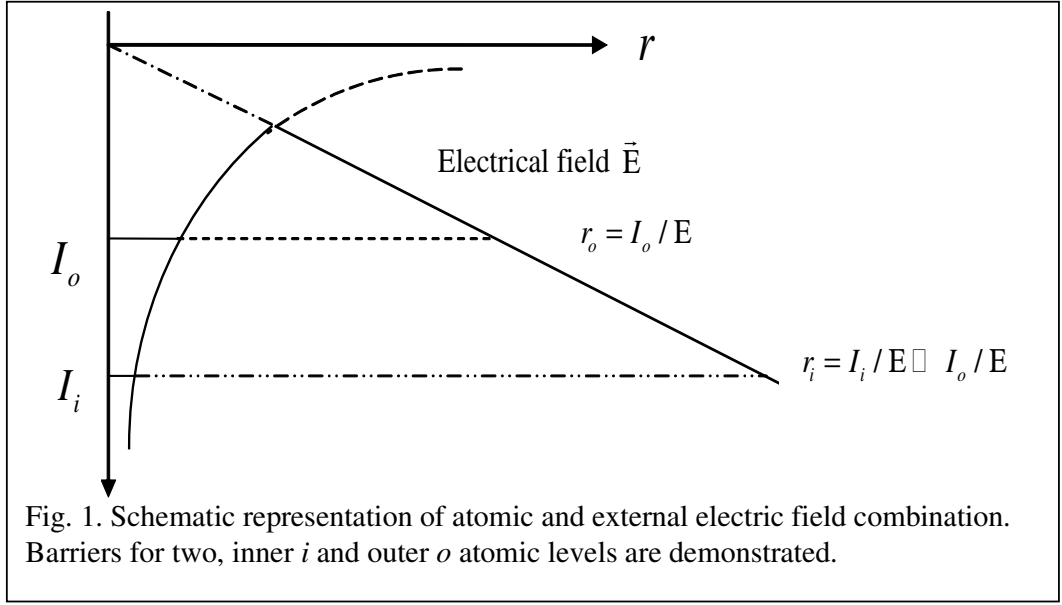

Fig. 1. Schematic representation of atomic and external electric field combination. Barriers for two, inner $i$ and outer $o$ atomic levels are demonstrated.

The probability to be ionized by the static field $\vec{E}$ for electrons $i$ and $o$ is determined by the probability to find corresponding electrons at points $r_i = I_i / E$ and $r_{n_o} = I_{n_o} \ll r_i$. As is well known, this probability is given by square modulus of the corresponding wave functions at points $r_i$ and $r_{n_o}$. Assuming that these points correspond already to the asymptotic region for the wave function, we receive in one-electron approximation for $i$ and $o$ electrons:

$$"i" \ |\varphi_i(r_i)|^2 \approx \alpha_i^3 (\alpha_i r_i)^{2\left(\frac{1}{\alpha_i}-1\right)} e^{-2\alpha_i r} \approx I_i^{\frac{1}{\alpha_i}+\frac{1}{2}} \left(\frac{I_i}{E}\right)^{2\left(\frac{1}{\alpha_i}-1\right)} e^{-2\sqrt{I_i}\frac{I_i}{E}}$$

$$"o" \ |\varphi_{n_o}(r_{n_o})|^2 \approx \alpha_{n_o l}^3 (\alpha_{n_o l} r_{n_o})^{2\left(\frac{1}{\alpha_{n_o l}}-1\right)} e^{-2\alpha_{n_o l} r} \approx I_i^{\frac{1}{\alpha_{n_o l}}+\frac{1}{2}} \left(\frac{I_{n_o l}}{E}\right)^{2\left(\frac{1}{\alpha_{n_o l}}-1\right)} e^{-2\sqrt{I_{n_o l}}\frac{I_{n_o l}}{E}}$$

(8)

Let us now introduce inter-electron Coulomb interaction and exchange. According to consideration in section 3, the wave function of inner electron is given by (5) that lead to another decay probability than (8), namely

$$"i" \ |\varphi_{i,ex}(r_i)|^2 \approx \left| \alpha_i^{3/2} (\alpha_i r_i)^{\frac{1}{\alpha_i}-1} e^{-\alpha_{il} r} + \frac{C_{n_o}}{(\alpha_i r_i)^2} \alpha_{n_o l}^{3/2} (\alpha_{n_o l} r_i)^{\frac{1}{\alpha_{n_o l}}-1} e^{-\alpha_{n_o l} r_i} \right|^2. \tag{9}$$

For deep levels the contribution of the first term is negligible, so that the penetration of the electron out of the atom is given by expression

$$|\varphi_{i,ex}(r_i)|^2 \approx \frac{\alpha_{n_o}^3 C_{n_o}^2}{(\alpha_i r_i)^2} (\alpha_{n_o l} r_i)^{\frac{2}{\alpha_{n_o l}}-2} e^{-2\alpha_{n_o l} r_i}. \tag{10}$$

The enhancement factor $\eta$ due to inclusion of the Fock term into the one-electron wave function of the inner electron is determined by the ratio of (10) to the expression in the first line in (8)

$$\eta \equiv \frac{|\varphi_{i,ex}(r_i)|^2}{|\varphi_i(r_i)|^2} = \frac{\alpha_{n_o l}^3 C_{n_o}^2}{(\alpha_i r_i)^2} (\alpha_{n_o} r_i)^{2\left(\frac{1}{\alpha_i}-\frac{1}{\alpha_{n_o l}}\right)} e^{2(\alpha_i - \alpha_{n_o l}) r_i} \approx$$

$$\approx C_{n_o}^2 \left(\sqrt{2 I_{n_o l}} I_i / E\right)^{2\left(\frac{1}{\alpha_i}-\frac{1}{\alpha_{n_o l}}\right)} \exp(2\sqrt{2 I_i} I_i / E) \tag{11}$$

It is remarkable that if there are $N_o$ outer electrons, the factor $\eta$ in accord with (7) acquire an additional enhancement factor $N_o^2$.

To illustrate the possible size of the enhancement factor $\eta$, let us consider a numerical example, in which the inner electron binding energy $I_i$ is five atomic units, while the outer electron binding energy $I_o$ is half atomic unit and the external field intensity E is one unit[3]. Then the factor $\eta$ is of the order of $5.64 \times 10^{13}$, while for the same field and levels energy one and ten atomic units, respectively one have $7.86 \times 10^{38}$!

These tremendously big numbers are a consequence of extremely small probability to eliminate an inner electron without exchange with the outer. Therefore it is more interesting and instructive to compare the ratio $\tau$ of inner to outer electron ionization probabilities when the exchange between outer and inner electrons is taken into account. This ratio is given by the following expression

---

[3] That corresponds to the field intensity $10^{16}$ W/cm$^2$.

$$\tau \equiv \frac{\left|\varphi_{i,ex}(r_i)\right|^2}{\left|\varphi_o(r_o)\right|^2} = \frac{\alpha_{n_o l}^3 C_{n_o}^2}{(\alpha_i r_i)^4} (\alpha_{n_o l} r_i / r_{n_o})^{\frac{2}{\alpha_{n_o l}}-2} e^{2\alpha_{n_o l}(r_{n_o}-r_i)} \approx$$
$$\approx \frac{\alpha_{n_o l}^3 C_{n_o}^2}{(\alpha_i r_i)^4} (\alpha_{n_o l} r_i / r_{n_o})^{\frac{2}{\alpha_{n_o l}}-2} e^{-2\sqrt{2I_{n_o l}} I_i / E} \sim E^4 \exp(-2\sqrt{2I_{n_o l}} I_i / E)$$
(12)

For the considered examples the energies of two levels, it is obtained for $\tau \approx 4.49 \times 10^{-5}$ and $5.01 \times 10^{-13}$. For the first case the ratio is not too small.

Qualitatively, it looks like the exchange admixture of outer electron literally "drags out" the inner electron off the ionized atom.

As it was mentioned before, if it is $N_o$ outer electrons, for which the coefficient $C_{n_o}$ is non-zero, this ratio is increased by additional factor $N_o^2$. It seems that this dependence was really observed in a number of investigations (see e.g. [12]) of multiple photoionization of noble-gas clusters by high intensity laser beam. In this studies a prominent amount of photons with energies of several hundreds of eV were found signaling the possibility that vacancies in inner shells were generated during laser-cluster interaction. The intensity of such processes in clusters could be a direct consequence of presence of very many outer electrons in clusters, contrary to the case of isolated atoms.

Note that exchange effects could be strongly enhanced even if the target atom exists in an exited state for a relatively short time. Therefore presence of strong atomic resonances, e.g. Giant, at laser frequency can enhance the multiple ionization probability considerably. Perhaps this is the reason why in photoionization by free-electron laser an abundance of multiply charged ions, with degree of ionization up to twenty-one were found [13].

It was demonstrated recently both numerically and analytically in [8, 9] that the Fock exchange leads in fact to non-exponential instead of exponential barrier penetration probability. This is seen qualitatively already from [(11)]: if $\alpha_{n_o l} r_o \sim 1$, the second term presents power decrease of the barrier penetration probability.

**5.** In this Letter we have presented a new mechanism that can explain considerable, of many orders of magnitude, increase of the inner electron ionization probability by a strong electric field. This increase is an inevitable result of exchange between the inner and outer electrons that dramatically change the inner electron wave function asymptotic.

The result presented here may be essential in other domains, where the HF approach successfully penetrated. Very often HF results are compared to that obtained in the frame of LDA – Local Density Approximation [14]. It is necessary to have in mind, however, that LDA by definition lacks non-locality. This is why in some respect the results of calculations using both these approaches can differ considerably.

As appropriate objects for HF equations, however much more difficult for calculations, are atoms imbedded in condensed matter objects, clusters or fullerenes. However, since they have much more outer electrons than an isolated atom, the wave function of an inner electron is modified stronger than in atoms.

One can expect traces of the discussed effects in atomic collisions in the strong laser field, while temporarily effectively exchanging objects are formed.

Similar effect could be of importance in objects that are described by two and one-dimensional HF equations.

HF equations [15, 16] were studied and applied mainly to multi-fermion systems. For bosons the exchange correction does not eliminate self-action, leading instead to doubling of its effect. However the whole concept of the role of non-locality could be of interest there also.

Giving the approximate nature of HF approach, it is essential to know whether account of electron correlations preserves of destroys the Fock's exchange contribution. It is possible to show using the example of infinite electron gas that non-locality is preserved but in case of high density electron gas noticeably diminished. In [9] arguments are presented that correlations in atoms correct the inner electron asymptotic but by terms of higher powers in $1/r$ than the Fock term.